\documentclass[%
 reprint, superscriptaddress,
 amsmath,amssymb, aps, prl,
]{revtex4-1}
\usepackage{color}
\usepackage{graphicx}
\usepackage{dcolumn}
\usepackage{bm}
\usepackage{dcolumn}
\usepackage{gensymb}
\usepackage{siunitx}
\newcommand{\RNum}[1]{\uppercase\expandafter{\romannumeral #1\relax}}

\begin{document}

\preprint{APS/123-QED}

\title{Machine Learning Energies of 2\,M Elpasolite (ABC$_2$D$_6$) Crystals}

\author{Felix Faber}%
\affiliation{Institute of Physical Chemistry and National Center for Computational Design and Discovery of Novel Materials, Department of Chemistry, University of Basel, 4056 Basel, Switzerland.}
\author{Alexander Lindmaa}%
\affiliation{Department of Physics, Chemistry and Biology, Link\"oping University, SE-581 83 Link\"oping, Sweden.}
\author{O. Anatole von Lilienfeld}%
\email{anatole.vonlilienfeld@unibas.ch}%
\affiliation{Institute of Physical Chemistry and National Center for Computational Design and Discovery of Novel Materials, Department of Chemistry, University of Basel, 4056 Basel, Switzerland.}
\affiliation{General Chemistry, Free University of Brussels, Pleinlaan 2, 1050 Brussels, Belgium}
\author{Rickard Armiento}%
\email{rickard.armiento@liu.se}%
\affiliation{Department of Physics, Chemistry and Biology, Link\"oping University, SE-581 83 Link\"oping, Sweden.}
 

\date{\today}

\begin{abstract}

Elpasolite is the predominant quaternary crystal structure (AlNaK$_2$F$_6$ prototype) 
reported in the Inorganic Crystal Structure Database. 
We have developed a machine learning model to calculate density functional theory quality formation energies of all $\sim$2\,M  pristine ABC$_2$D$_6$ elpasolite crystals which can be made up from main-group elements (up to bismuth). 
Our model's accuracy can be improved systematically, reaching 0.1 eV/atom for a training set consisting of 10\,k crystals. 
Important bonding trends are revealed, fluoride is best suited to fit the coordination of the D site which lowers the formation energy whereas the opposite is found for carbon. 
The bonding contribution of elements A and B is very small on average.
Low formation energies result from A and B being late elements from group (II), 
C being a late (I) element, and D being fluoride.
{\color{black} Out of 2\,M crystals, 
{\color{black} 90 unique} structures are predicted to be on the convex hull---among which NFAl$_2$Ca$_6$, 
with peculiar stoichiometry and a negative atomic oxidation state for Al.
}

\end{abstract}

\maketitle



\begin{figure}
\centering
\includegraphics[width=1.\linewidth]{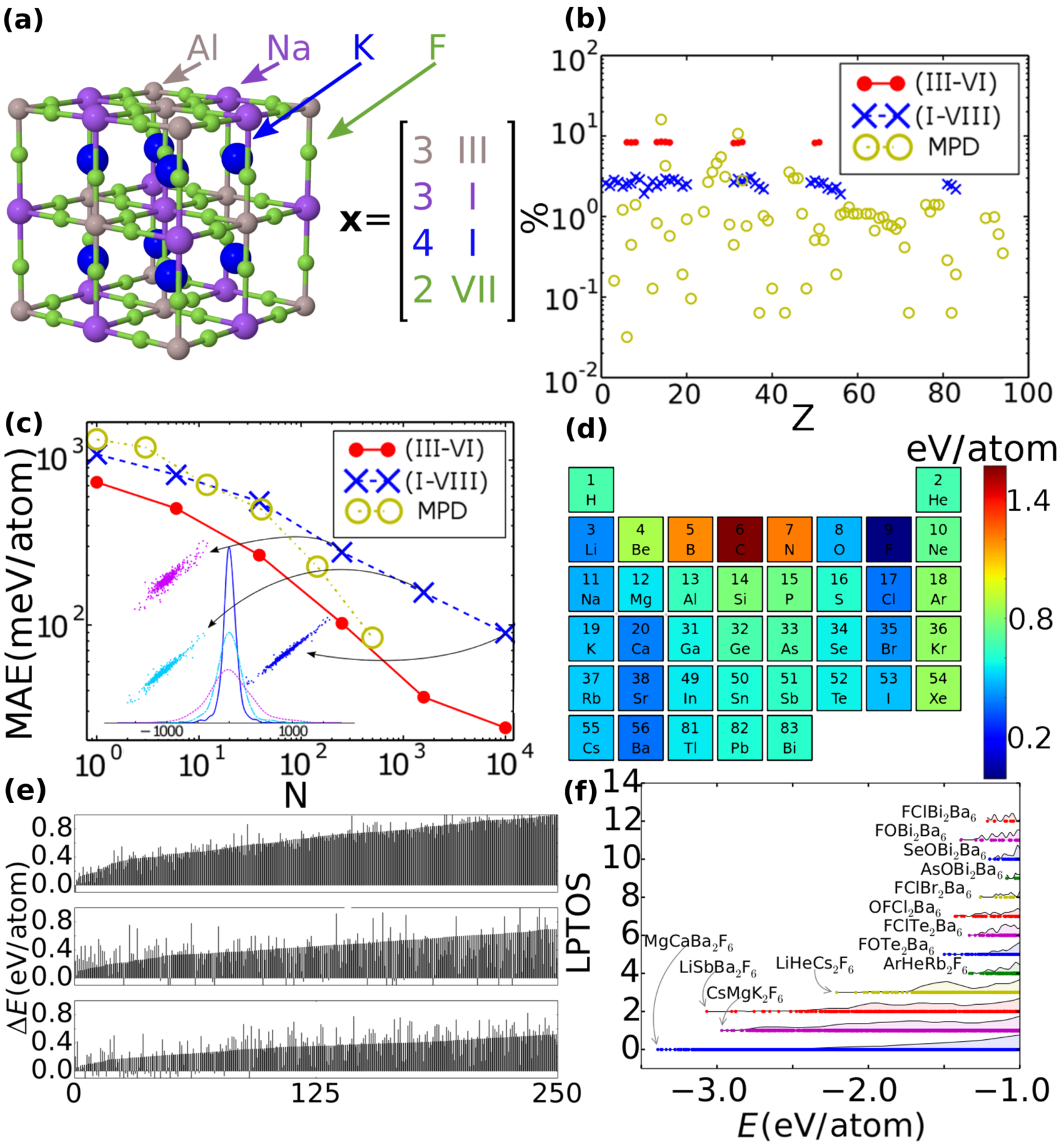}
\caption{
\label{fig:ElementDistribution}
\label{fig:NewOrder}
\label{fig:ErrorVsSizeAndErrDist}
\label{fig:scheme}
(a) Illustration of elpasolite crystal (AlNaK$_2$F$_6$ structure). 
The four-tuple $x = (x_1,\ldots,x_4)$ representation of atomic sites is specified. 
(b) Frequency of elements (defined by nuclear charge Z) for the three data sets studied. 
(c) Mean absolute out-of-sample prediction error as a function of training set size for the three data sets studied. 
Inset: Error distributions and DFT vs.~ML scatter plots for three training set sizes for the (\RNum{1}$-$\RNum{8}) data set.  
(d) Estimated mean energy contribution of each element to formation of any elpasolite crystal. 
    The color code reflects the new elemental elpasolite order.
(e) Lowest 250 ML model predicted formation energies of elpasolites in ascending order from (\RNum{3}$-$\RNum{6}) (TOP) 
and (\RNum{1}$-$\RNum{8}) (MID and BOTTOM) data sets.
Results in TOP and MID panel correspond to ML models trained on 2000 examples, BOTTOM panel results 
correspond to a ML model trained on 10k crystals.
Validating DFT energies are shown aside.
(f) Distributions of absolute lowest possible total oxidation states (LPTOS) in energies. 
Formulas indicate the lowest lying crystals. 
}
\end{figure}

Elpasolite (AlNaK$_2$F$_6$) is a glassy, transparent, luster, 
colorless, and soft quaternary crystal in the Fm3m space group which can
be found in the Rocky Mountains, Virginia, or the Apennines. The elpasolite crystal structure (See  Fig.~\ref{fig:scheme}) is not uncommon, 
it is the most abundant prototype in the Inorganic Crystal Structure Database \cite{icsd,icsd2}.
Some elpasolites emit light when exposed to ionic radiation, which makes them interesting material candidates for scintillator devices \cite{ElpasoliteHalide,BiswasElpasolite}. 
One could use first-principle methods such as density functional theory (DFT)~\cite{DFT1,DFT2} 
to computationally predict the existence and basic properties of every elpasolite. 
Unfortunately, even when considering crystals composed of only main group elements (columns I to VIII) the sheer number of the $\sim$2\,M possible combinations makes DFT based screening challenging---if not prohibitive. 
Recently, computationally efficient machine learning (ML) models were introduced for predicting molecular properties with the same accuracy as DFT~\cite{M.Rupp,Montavon2013}. 
Requiring only milliseconds per prediction, they represent an attractive alternative when it comes to the combinatorial screening of millions of crystals. 
While some ML model variants have already been proposed for solids \cite{Gross,Meredig_2014,QUA:QUA24917},
a generally applicable ML-scheme with DFT accuracy of formation energies is still amiss.

In this Letter we introduce a newly developed ML model which we use to investigate the 
formation energies of \emph{all} $\sim$2\,M elpasolites made from all main-group elements up to Bi.
{\color{black}
Resulting estimates enable the identification of new elemental order of descending elpasolite formation energy, 
crystals with peculiar atomic charges, 
250 elpasolites with lowest formation energies, 
as well as 128 new crystal structures predicted to lie on the convex {\color{black} hull}
among which 
NFAl$_2$Ca$_6$, an elpasolite with unusual composition and atomic charge. 
}
The ML model achieves {\color{black} the same, or better, accuracy to DFT as DFT to experimental data} and can be generalized to any crystalline material.


The ML-model is based on kernel ridge regression \cite{muller2001introduction,scholkopf2002learning,kernel_ridge_regression2} 
which maps the non-linear energy difference between the actual DFT energy
and an inexpensive approximate baseline model into a linear feature space~\cite{Delta-learning}. 
More specifically, we construct a ML model of the energy difference to the sum 
of static, atom-type dependent, atomic energy contributions $\epsilon_{It}$,
obtained through fitting of each atom type $t$ in all main group elements up to Bi. 
The energy-predicting function is a sum of weighted exponentials in similarity $d$ between query and training crystal, 
\begin{equation}
E(\mathbf{x}) = \sum_I^{N'} \epsilon_{It} + \sum_{i }^{N} \alpha_i e^{-d_i/\sigma},
\end{equation}
where $N'$ is the number of atoms/unit cell (10 in the case of elpasolites), and the second sum runs over all $N$ training instances. 
$\alpha_i$ are the weights obtained through linear regression, and
$\sigma$ is the global exponential width, regulating the length scale of the problem.  
The similarity $d_i$ is the Manhattan distance, i.e., $d_i = \left \| \mathbf{x} - \mathbf{x}_i  \right \|_1$. 
While various crystal structure representations have previously been proposed~\cite{Gross,Meredig_2014,QUA:QUA24917,LucaDescriptor2015},
we have found the following representation to yield superior performance:
$\mathbf{x}$ is a $n\times 2$ tuple that encodes any stoichiometry within a given crystal prototype. 
For quaternary ($n = 4$) elpasolites, each $x_{1-4}$ refers to the 4 representative sites,
the atom type for each site is represented by its row (principal quantum number 2 to 6) and column (number of valence electrons) I to VIII in the periodic table,
and sites are ordered according to the Wyckoff sequence of the crystal.
As such, $\mathbf{x}$ implicitly represents the energy minimum structure for a system restricted to this prototype---without explicitly encoding precise coordinates, lattice constants, or other (approximate) solutions to Schr\"odinger's equation.
This representation is not restricted to the elpasolite structure, it can be used for any crystalline configuration: Below we also briefly discuss test results for small size ML models applied to ternary crystals.

For training and evaluation, we have generated DFT 
{\color{black} 
formation energies for two data sets of elpasolites 
(for computational details see supplementary materials, Ref.~\onlinecite{supplementary}),
one small, (\RNum{3}$-$\RNum{6}), made up from only 12 elements, C, N, O, Al, Si, P, S, Ga, Ge, As, Sn, and Sb;
and one large, (\RNum{1}$-$\RNum{8}), containing all main-group elements up to Bi.
Since (\RNum{3}$-$\RNum{6}) only comprise $\sim$12\,k possible permutations, 
we have obtained the complete list of formation energies. 
}
(\RNum{1}$-$\RNum{8}) consists of 10\,k structures, i.e.~0.5\% of the total number of 2\,M possible crystals. 
The (\RNum{1}$-$\RNum{8}) data set has been generated through random selection of 
elpasolites while ensuring an unbiased composition. 
To verify that the ML model is general and not only restricted to elpasolites, 
we have also included a materials project \cite{materals_project1} dataset (MPD)
consisting of $\sim$0.5k ternary crystals in $\mathrm{ThCr}_2\mathrm{Si}_2$ (I4/mmm) prototype and
made up of 84 different atom types. 
The distribution of the chemical elements in the data sets are shown in Fig.~\ref{fig:ElementDistribution}(b). 
Numerical results on display in Fig.~\ref{fig:ErrorVsSizeAndErrDist}(c) indicate systematic improvement of  
the predictive accuracy of the ML model with increasing training set size, for all three datasets. 
The inset details normally distributed errors and scatter plots which systematically improve with training set size for the models trained on the (\RNum{1}$-$\RNum{8}) datasets. 
{\color{black} For a 10k training set, the ML model reaches a MAE of 0.1 eV/atom compared to reference, i.e.~semi-local DFT.
DFT, in turn, has an estimated MAE of $\sim$0.19 eV/atom compared to experiments on heats of formation for general chemistries
with filled $d$-shells~\cite{PhysRevB.78.245207}. 
For transition metal oxides and elemental solids other groups report DFT errors on the order of 0.1 eV/atom~\cite{Ternary_oxides,AnnRickardSolids2008jcp}}.
The converging performance for training on nearly all crystals of the (\RNum{3}$-$\RNum{6}) data set 
suggests that our {\color{black} crystal representation of elpasolite structures Fig.~\ref{fig:ErrorVsSizeAndErrDist}(a) accounts for all necessary degrees of freedom. 
While errors decay systematically and linearly on a log-log plot, the learning rate levels off 
as $N$ approaches the 100\%, i.e.~$10$k. 
This is due to the employed relaxation convergence threshold of $\pm 10$ meV/atom in the DFT calculations.
Any inductive model must fail to go below this level, and only numerically more precise reference numbers would mitigate this trend.}
In all validation tests dealing with energy predictions for random out-of-sample crystals, 
the ML model performance meets the expectations set in Fig.~\ref{fig:ErrorVsSizeAndErrDist}(c). 
For example, drawing 100 crystals at random from (\RNum{3}$-$\RNum{6}) and (\RNum{1}$-$\RNum{8}) datasets
ML models perform as expected when compared to the result from validating DFT calculations (cf.~Fig.~S3~\cite{supplementary}). 
{\color{black} (\RNum{3}$-$\RNum{6}) and (\RNum{1}$-$\RNum{8}) reaches a MAE of 0.1 eV/atom at roughly 2.5 \% and 0.5 \% of the total number of crystals respectively, suggesting that the machine "efficiency" increases with number of possible combinations. We note however that two observations of the same structure is not sufficient to see any trends on how much training data is needed.}

\begin{figure*}
{\centering
  \includegraphics[width=0.8\textwidth]{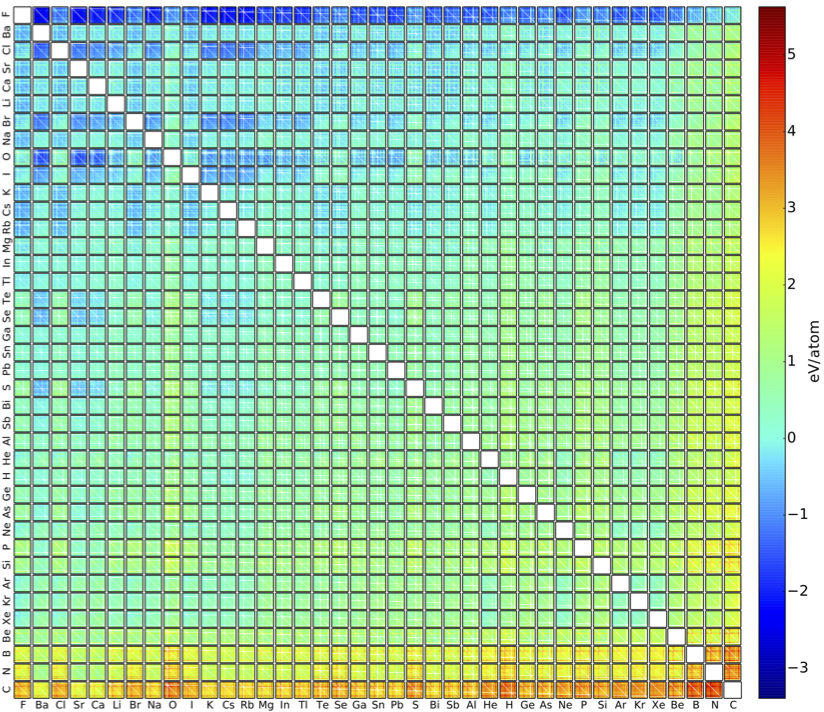}\\
  }
\caption{
\label{fig:AllEnergies-c}
Formation energies for all 2\,M elpasolites made up of all main-group elements up to Bi predicted by the 10\,k ML-model. 
The outer vertical and horizontal axis correspond to $x_4$ and $x_3$ symmetry position, respectively.
Inner vertical and horizontal axis correspond to $x_2$ and $x_1$ symmetry position, respectively. 
Elemental sequence follows the elpasolite order of Fig.~\ref{fig:ErrorVsSizeAndErrDist}(d).
White pixels correspond to subspaces of ternary, binary, or elementary non-elpasolite crystals.}
\end{figure*}

Having established the performance of the ML model, we have subsequently used the 
10 k training set model (\RNum{1}$-$\RNum{8}) for investigation of the elpasolite universe.
Estimated formation energies for {\em all} 2\,M elpasolites are featured in 
Fig.~\ref{fig:AllEnergies-c}.
The formation energies are clearly dominated by the chemical identity of position 4, 
followed by position 3 but according to a different pattern. 
Chemical identity at position 1 and 2 has the smallest influence and very similar impact 
(also illustrated in Fig.~S1 of Ref.~\onlinecite{supplementary}.)
Due to the effective degeneracy of positions 1 and 2, all inner matrices in Fig.~\ref{fig:AllEnergies-c}
appear largely symmetric. 
Figure ~\ref{fig:ErrorVsSizeAndErrDist}(d) shows the average contribution of each element to the formation energies estimated by the 10\,k ML model. These average contributions per element are used to order the elements in Fig.~\ref{fig:AllEnergies-c} to yield the smoothest elpasolite map. Arranging elements by their nuclear charge, or by their Pettifor order~\cite{D.Pettifor}, results in a much more oscillatory map or stripe-like pattern due to underlying periodicities (cf.~Ref.~\onlinecite{supplementary}). 
{\color{black}
This elpasolite error is dominated by the element identity in position 4 (compare Figure ~\ref{fig:ErrorVsSizeAndErrDist}(d) to Fig.~S1 of Ref.~\onlinecite{supplementary}); 
its break-down is small as illustrated for pair-wise energy contributions in Fig.~S5 of Ref.~\onlinecite{supplementary}.
}

Figures~\ref{fig:ErrorVsSizeAndErrDist}(d) visualize the bonding emergent from the geometry and bond coordination of the elpasolite crystal structure (see also figures in supplementary materials). 
Fluorine and carbon are at the respective ends of the global scale of low and high formation energies. But also alkaline metals, alkaline earth metals, and oxygen contribute to lowering the formation energy. 
On average, the formation energies of elpasolites involving halogens, alkaline metals, noble gases increase as the periodic table is descended. 
The opposite holds for all other elements, except oxygen, boron, carbon and nitrogen, 
which all have a noticeably higher average formation energy than any other element. 
A saddle point can also be observed in the midst of the periodic table table as well as two valleys along the halogen and alkaline earth rows.
Site-specific resolution indicates that fluorine fits best with the bond coordination of sites 1, 2, and 4, whereas the same does not apply to later halogens (not shown in the paper, see supplementary materials~\onlinecite{supplementary}). 
In contrast, as the element on site 3 goes down column II in the periodic table, the formation energy is successively lowered, with Ca, Sr, and Ba contributing more than any halogen atom. On sites 1 and 2, the formation energy generally increases the most for heavy noble gases. On sites 3 and 4, it is carbon, followed by neighboring B and N that increase the formation energy the most. The accuracy of linear single atom energy models based on these scales, however, is not on par with the ML-model, and---maybe more importantly---cannot be improved systematically through increasing training set sizes but rather converges to a finite residual error.


In order to achieve satisfying accuracy of $\pm$0.1 eV/atom for elpasolites, a relatively large training set of 10\,k is needed. 
This is likely due to the sparsity of crystals at the opposite ends of the high and low formation energy spectrum; 
this results in a decreased predictive ML model accuracy for crystals in these regions{\color{black}, which is demonstrated in Fig S6 in Ref.~\onlinecite{supplementary}}.
Nevertheless, the 10\,k ML model readily identifies a larger set of lowest lying elpasolites for which the actual DFT minima can be obtained through subsequent DFT based screening. 
This is shown in Fig.~\ref{fig:ErrorVsSizeAndErrDist}(e) where the 250 crystals with the lowest ML predicted formation energies are shown in ascending order (with further details on these systems in the supplementary mateirals.~\onlinecite{supplementary}.) Subsequent screening with DFT indicates the 26$^{th}$ crystal CaSrCs$_2$F$_6$ (out of 2M) to be the global formation energy minimum at $-3.44$ eV/atom, closely followed a near-degenerate isomer SrCaCs$_2$F$_6$. The DFT energies of the next two degenerate pairs CaSrRb$_2$F$_6$/SrCaRb$_2$F$_6$ and CaBaCs$_2$F$_6$/BaCaCs$_2$F$_6$ 
correspond to $-3.41$, and $-3.39$ eV/atom, respectively.
Overall, the elpasolites with the most favorable formation energies, ABC$_2$D$_6$, correspond to A and B being late elements from group (II),
and C and D being a late element from group (I) and fluoride, respectively.
Populating the four sites with elements from groups (II),(II),(I), and (VIII), respectively, differs from the experimentally 
established stoichiometry AlNaK$_2$F$_6$. 
In fact, the lowest DFT energy crystal with a group-(III) element is CsAlRb$_{2}$F$_{6}$ (in 69$^{th}$ position) with
$-3.09$ eV/atom (ML energy: $-2.96$ eV/atom, see supplementary materials).

We have also used our predictions to analyse atomic oxidation states in elpasolites. 
In particular, we have found that roughly 6 \% of the crystals with formation energies below $-1$ eV/atom exhibit unusual atomic charges: 
They are low in energy despite the fact that no combination of conventional atomic charges would result in a neutral system. 
In order to identify these crystals, we have used the absolute value of the lowest possible total oxidation state (LPTOS) 
that could possibly be realized using the list of typical atomic oxidation states on display in 
{\color{black} Table~III
in Ref.~\onlinecite{supplementary}.}
The lowest lying crystals have a LPTOS of $0$ ($-3$ to $-3.44$ eV/atom formation energies). 
However, already at $-3$ eV/atom crystals with LPTOS of 2 or 1 start to occur. 
At formation energies of $\sim-1.25$ eV/atom and higher, the number of crystals with non-zero LPTOS increases rapidly, 
with LPTOS as high as $12$.
Corresponding crystal frequency distributions are shown in Fig.~\ref{fig:ErrorVsSizeAndErrDist}(e), 
along with formulas for the mutually lowest lying crystals.
Interestingly, the number of crystals with zero LPTOS increases monotonically with formation energy, 
while for nonzero LPTOS crystals the distribution is oscillatory.

{\color{black} 
 To demonstrate the usefulness of our ML model we have applied it to identify thermodynamically stable elpasolites.
 To this end, we first selected all those 274,213 elpasolites with negative ML formation energies, and without rare gas elements. 
 Since stability depends on the energy difference to any possible polymorph or competing segregated phases \cite{ADMA:ADMA200700843, Ong20081798}, 
 we have queried available DFT formation energies stored in the Materials Project (MP)~\cite{materals_project1}.  
{\color{black}  
 Some elpasolites, such as the archetypical AlNaK$_2$F$_6$, are already stored in the MP.  
 Using DFT results stored in the MP for competing quaternary, ternary and binary phases, 
 we have constructed phase diagrams~\cite{Ong20081798} for all 274,213 crystals.
 For each crystal, there are on average $\sim$12 competing phases stored in the MP; there is only one combination of elements (Cs-Li-Na-Rb)
for which no binary or ternary competing phases have been stored.}
 For 2133 out of the 274,213 crystals, the resulting stabilization energy is below the known convex hull of stability (for more details, see Ref.~\onlinecite{supplementary}).
 Subsequent validation using DFT instead of ML confirms 128 out of them to be stable. Out of these 38 are polymorphs (ABC$_2$D$_6$ vs.~BAC$_2$D$_6$)
 resulting in 90 unique stoichiometries.  
 Such a reduction (274,213 $\rightarrow$ 90) in number of crystal candidates is to be expected since sorting crystals by ML energies being lower than the convex hull 
 systematically favors those with negative ML formation energy errors.
 We note that this does not amount to proof that the {\color{black}90} crystals are stable: 
 The MP database is not exhaustive. This implies that other new competing phases and materials, 
 with even stronger stabilization, might still be discovered in the future.  
 Also, the intrinsic error of the employed DFT method within the MP might still alter the outcome with respect to experiment. 
As such, the {\color{black} 90} new elpasolite DFT energies represent new upper bounds on the convex hull at the corresponding compositions. 
They have been submitted to the MP database, and most of them have been made available for further studies 
(See Table V in Ref.~\onlinecite{supplementary} for the list of the 90 structures).

 Among these elpasolites, metals, semiconductors and insulators are roughly distributed equally.
 All structures with an earth alkaline metal in crystal position 4 have a low or zero band-gap. 
We have noted an intriguing yet stable structure of a conductor, NFAl$_2$Ca$_6$ (MP ID: mp-989399, \# {\color{black} 20} in Table V in Ref.~\onlinecite{supplementary}) with Ca at position 4, instead of F or Cl. 
Bader charge analysis~\cite{bader1,bader2,bader3} (Table~\ref{tab:NFcharge}) 
indicates an exotic negative oxidation state for Al (-II), 
previously only reported for Al in substantially larger Zintl phase unit cells (Sr$_{14}$[Al$_4$]$_2$Ge$_3$)~\cite{AlGeSr}. 
Since Bader charges sometimes yield non intuitive results~\cite{Voronoi1,1367-2630-15-11-115007},
calculated Hirshfeld~\cite{Hirshfeld} and Voronoi deformation density~\cite{Voronoi1,Voronoi2} 
charges (Table \ref{tab:NFcharge}) confirm the negative oxidation state, albeit reduced by one unit (-I). 
The calculated phonon spectra of NFAl$_2$Ca$_6$ also indicate stability [\onlinecite{supplementary}]. 
}

\begin{table}
\caption{
\label{tab:NFcharge}
{\color{black}
Calculated atomic charges in NFAl$_{2}$Ca$_{6}$ elpasolite using different methods
(obtained using {\tt SIESTA}\cite{soler2002siesta}).
}
}
\begin{tabular}{l  
S[table-number-alignment = center] 
S[table-number-alignment = center]
S[table-number-alignment = center]
S[table-number-alignment = center]}
 Method & N &  F  & Al & Ca \\
 \hline
Bader & -2.00 & -0.98 &  -2.13 & 1.20 \\
Hirshfeld & -0.63 & -0.36 &  -1.05 & 0.52 \\
Voronoi deformation density & -0.81 & -0.29 & -1.13 & 0.56 \\
\end{tabular}
\end{table}


In conclusion, we have developed and used ML-models of formation energies to investigate all possible elpasolites made up of main-group elements. 
We have presented numerical results for $\sim$2 M formation energies. 
The ML-model is only implicitly dependent on spatial coordinates, through reference data used for training. 
No spatial coordinates are needed for new queries, yet for a training set of 10\,k crystals the model reaches $\pm$0.1 eV/atom---comparable to DFT accuracy for solids. 
The results have been used to identify the most strongly bound elpasolites as well as to 
investigate energy and bonding trends at crystal structure sites, leading to a new ``elpasolite order'' of elements, consistent with the bonding physics in the elpasolite crystal structure.
{\color{black}
We identified and added 128 structures {\color{black}(90 unique stoichiometries)} to the convex hull of the MP database. 
Charge analysis for the metallic elpasolite NFAl$_2$Ca$_6$ indicates a negative atomic oxidation state of Al. 
This outcome directly demonstrates that our method can be used for the discovery of stable as well as unconventional chemistries. 
Due to the low computational cost of the ML model one can now also afford to remove human bias by considering also those structures 
which previously would have been excluded due to ''chemical intuition''.
Our results suggest that ML models hold} great promise for the computational screening of polymorphs, other crystal structure symmetries, solid mixtures, phase transitions, 
or defects at unprecedented rate and extent. 
Other crystal properties than energies could also be considered.

\begin{acknowledgements} 
The authors thank G.~Hart, R.~Ramakrishnan and R.~Ramprasad for comments; A.~Jain, D.~Winston, P.~Huck and K.~A.~Persson for helpful 
input on stability calculations and for validating elpasolites for inclusion in MP; and B.~Huang and R.~Sarmiento-P\'erez for help 
with further validation of stability and the charges of NFAl$_2$Ca$_6$. 
O.A.v.L. acknowledges funding from the Swiss National Science foundation (No.~PP00P2\_138932). 
This material is based upon work supported by the Air Force Office of Scientific Research, Air Force Material Command, USAF under Award No.~FA9550-15-1-0026. 
This research was partly supported by the NCCR MARVEL, funded by the Swiss National Science Foundation.
This work was supported by a grant from the Swiss National Supercomputing Centre (CSCS) under project ID mr14. 
R.A. acknowledges funding by the Swedish Research Council Grant No. 621-2011-4249 and Linnaeus Environment grant (LiLi-NFM). 
Calculations have been performed at the Swedish National Infrastructure for Computing (SNIC).
\end{acknowledgements} 

\bibliographystyle{apsrev4-1}
\bibliography{ref}
\end{document}